\documentclass[3p,twocolumn]{elsarticle}
\usepackage{amssymb}
\usepackage{epsfig}
\usepackage{pifont}
\usepackage{booktabs}

\journal{Astroparticle Physics}

\usepackage{todonotes}
\let\oldtodo\todo
\renewcommand{\todo}[1]{%
	\oldtodo[inline]{#1}%
}

\newsavebox\CBox
\newcommand\hcancel[2][0.5pt]{%
  \ifmmode\sbox\CBox{$#2$}\else\sbox\CBox{#2}\fi%
  \makebox[0pt][l]{\usebox\CBox}%
  \rule[0.5\ht\CBox-#1/2]{\wd\CBox}{#1}}

\usepackage{soul}

\begin{document}

\begin{frontmatter}
\title{A Run-Wise Simulation and Analysis Framework for Imaging Atmospheric Cherenkov Telescope Arrays} 

\author[uibk]{M.~Holler}
\ead{markus.holler@uibk.ac.at}

\author[lpnhe]{J.-P.~Lenain}
\author[llr]{M.~de~Naurois}
\author[uibk]{R.~Rauth}
\author[lapp]{D.\,A.~Sanchez}

\address[uibk]{Institut f\"ur Astro- und Teilchenphysik, Leopold-Franzens-Universit\"at Innsbruck, A-6020 Innsbruck, Austria}
\address[lpnhe]{Sorbonne Universit\'e, Universit\'e Paris Diderot, Sorbonne Paris Cit\'e, CNRS/IN2P3, Laboratoire de Physique Nucl\'eaire et de Hautes Energies, LPNHE, 4 Place Jussieu, F-75252 Paris, France}
\address[llr]{Laboratoire Leprince-Ringuet, Ecole Polytechnique, CNRS/IN2P3, F-91228 Palaiseau, France}
\address[lapp]{Univ. Grenoble Alpes, Univ. Savoie Mont Blanc, CNRS, LAPP, 74000 Annecy, France}
\begin{abstract}

We introduce a new simulation and analysis paradigm for Imaging Atmospheric Cherenkov Telescope (IACT) arrays, simulating the actual observation conditions as well as individual telescope configuration for each observation unit. Compared to existing frameworks, where simulations are usually generated using pre-defined settings, this \textit{run}-wise simulation approach implies more realistic simulations and hence reduced systematic uncertainties. The computational effort of this dedicated simulation concept is notably independent of the amount of different observation configurations but just scales linearly with observation time. This corresponds to a large advantage for increasingly complex current and future IACT arrays where the size of the phase space makes it computationally unfeasible to generate simulations that reach the requirements regarding systematics using the classical simulation scheme. 

\end{abstract}

\begin{keyword}
IACT \sep VHE Gamma-ray Astronomy \sep Simulation \sep Analysis Techniques 

\end{keyword}

\end{frontmatter}

\section{Introduction}
\label{sec_intro}

In the past years, the relatively recent field of ground-based gamma-ray astronomy using Imaging Atmospheric Cherenkov Telescopes (IACTs) has undergone several improvements. Instrument upgrades both enlarge the accessible energy range and improve the overall data quality of the existing instruments (\cite{2016_HessUpgrade}, \cite{2011_VeritasUpgrade}, \cite{2016_MagicUpgrade}). As for the data analysis, advanced photon reconstruction and analysis techniques led to improved reconstruction precision and event classification (\cite{2009_Model}, \cite{2014_ImPACT}, \cite{2009_TMVA}).

Given that in the case of IACTs the atmosphere corresponds to an integral part of the detector, a direct and continuous calibration is not possible. This emphasises the need for realistic Monte-Carlo (MC) simulations to obtain the instrument response for a given observation. Unlike the aforementioned advancements, the overall MC simulation and analysis strategy has up to now remained unchanged. The principle of this approach is similar for all major current IACT experiments (see, e.g., \cite{2006_HessCrab}, \cite{2016_MAGICCrab}), and it is currently also planned to be used for the future Cherenkov Telescope Array \cite{2013_CTA_MC}. In general, MC simulations are carried out for predefined and fixed observation and instrument settings, covering the needed range of important parameters. Those include, for instance, the assumed atmospheric profile, the simulated zenith and azimuth angle of the observation, the number of participating IACTs, as well as the configuration of the IACT cameras. For those parameters which are assumed to have a major influence on the observation, simulations with different parameter values are carried out. Instrument Response Functions (IRFs), i.e. properties that describe the behaviour of the detector, such as its energy resolution or effective detection area, are generated for the different simulations and usually stored in lookup tables. In case discrete parameter values were simulated, the IRFs of the adjacent grid points of the lookup table are either interpolated, or the IRF of the closest grid point is taken to obtain the response for a given observation.
While this approach has been proven to work reliably, it has several shortcomings. To reduce the systematic uncertainties of a given parameter $i$ as induced by the lookup approach down to an acceptable level, it is necessary to simulate a sufficient amount of grid points $n_i$. The interpolation helps to achieve a better estimate between the grid points, but is in most cases just a coarse simplification, implying unavoidable systematic uncertainties due to the methodology. The number of individual simulation sets to be generated in total is given as $N = \Pi_{i}\, n_i$, which implies a considerable increase of computation time when deciding to simulate more than just one configuration for a certain parameter. Consequently, a large fraction of the simulation configuration is fixed, implicitly assuming that their influence is minor. 
The whole problem is augmented for new and upgraded arrays with different IACT sizes, types, and camera configurations, making it increasingly difficult to hold the relative contribution of systematic uncertainties sufficiently low with respect to the nominal performance gain.

Here we introduce a new simulation and analysis framework as an alternative to the existing approach. It takes into account the actual observation conditions as well as individual telescope configurations of each observation unit of a given data set, which brings many improvements. Firstly, since the simulations are specific to the observations it is not necessary to simulate for many combinations of detector configurations. 
Such a direct connection between data and simulation also ensures an efficient usage of computational resources which scale linearly with observation time and are independent of the complexity of the system to be simulated. 
Additionally, generating a dedicated simulation for every observation with a great level of complexity directly implies a reduction of systematic uncertainties for all analysis-related aspects which rely on simulations. Here it is important to note that in ground-based gamma-ray astronomy the background level is typically derived directly from observations instead of simulations, implying that the corresponding systematic uncertainties are unchanged when switching the simulation approach. The new framework as described here is already fully implemented in the software of the High Energy Stereoscopic System (H.E.S.S.) \citep{2006_HessCrab}, an array of four identical IACTs (CT1-4) with $107\,\mathrm{m}^2$ effective mirror area and a fifth one (CT5) with $614\,\mathrm{m}^2$.

The general principles as well as the technical implementation of the new method are described in Section~\ref{sec_impl}, followed by an evaluation of the performance of the approach in Section~\ref{sec_perf}. Conclusions are drawn in Section~\ref{sec_conc}.

\section{Principles and Implementation}
\label{sec_impl}

The main motivation for the new approach was to have simulations that are as close as possible to the observational reality in terms of both observation conditions and detector configuration. In IACT gamma-ray astronomy, many observation conditions can in general be assumed to be constant for the time scales of a unit of continuous data taking. As for H.E.S.S., such an observation unit lasts up to $28\,\mathrm{min}$ and is called \textit{observation run}, or just \textit{run} \citep{2006_HessCrab}. For the method presented here, dedicated simulation sets are generated on a \textit{run}-by-\textit{run} basis, therefore it is henceforth called the Run-Wise-Simulation (RWS) concept. While this simulation approach is new in ground-based gamma-ray astronomy, it is common and standard in other fields, such as, e.g., high-energy particle physics \cite{geant4_2003}.

\subsection{Technical Framework}
\label{sec_techframe}

The RWS approach has been fully implemented in one of the two simulation and analysis chains in operation in H.E.S.S., named \textit{parisanalysis}. 
\begin{figure*}[t]
\includegraphics[width=\textwidth]{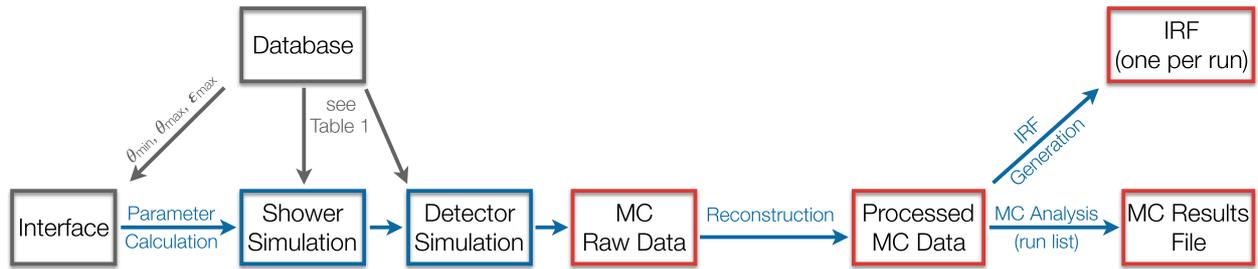}%
\caption{Overview flowchart of the RWS framework.}
\label{Fig_Flowchart}
\end{figure*}
A general overview of the implemented approach is shown in Fig.~\ref{Fig_Flowchart}. All information that is relevant for the simulation of a specific \textit{run} is stored in a MySQL database (DB) and read out on demand. The values stored in the DB mainly have two origins: the data-taking schedule and monitoring, directly obtained from the Data Acquisition system, and the calibration and analysis chains, performed offline.
Hence no read-in of \textit{run}-specific files is necessary, allowing the RWS framework to be easily ported and executed in distributed environments such as the European Grid Infrastructure (EGI).
The shower simulations are carried out with the KASKADE software \citep{1994_Kaskade}, which has been improved and rewritten in C++ to allow more flexible usage and easy integration in the simulation chain \citep{2009_Model}. KASKADE has been shown to provide results that are compatible with the ones of the widely used alternative simulation package CORSIKA \cite{1998_Corsika}, with differences regarding simulated Cherenkov photon distributions at the level of around $5\%$ \cite{2003_Thesis_Guy,2013_CTA_MC}. The code has been further enhanced for the RWS method to be able to simulate the evolution of the tracking position of the telescopes within a \textit{run}. The KASKADE output is directly passed to the internal IACT simulation software (see also Fig.~\ref{Fig_Flowchart}). No direct output of the shower simulation is saved in case of the implemented RWS approach, leading to a drastic reduction of needed disk space at the expense of increased computation time. To allow simulating larger statistics, simulations can be split into several computing processes for each \textit{run}. The simulated MC raw-data files, written in a format identical to that of real data, contain all information that is needed for the event reconstruction and higher-level analysis modules. Additionally, an auxiliary file is created for every \textit{run}, containing further details about the simulation configuration. A MySQL DB table is used to keep track of the simulation production, including details of the actual simulation parameters of each \textit{run}.

The produced MC raw-data are passed on to the event reconstruction, which is in the presented implementation the \textit{Model} analysis from \cite{2009_Model}. The reconstruction is conducted using the same routine as for measured raw data. The processed RWS MC data is suitable to be used for, e.g., performing a MC analysis of a set of \textit{runs} (Sec.~\ref{mc_ana}) or the generation of IRFs (Sec.~\ref{spectra}).

\subsection{Simulation Parameter Settings}
\label{sec_simpar}

Observations of IACT arrays generally cover a large range of observation conditions, implying substantial variations of the required simulation parameter space. For an efficient simulation with yet sufficient and consistent parameter coverage for each \textit{run}, an automated parameter calculation approach is indispensable. The relevant inputs for the calculation presented here are:
\begin{itemize}
\item List of active IACTs,
\item Range of zenith angles $\left[ \theta_{\mathrm{min}}, \theta_{\mathrm{max}} \right]$ covered within the \textit{run},
\item Optical efficiencies $\varepsilon$ of the participating IACTs relative to a non-degraded system,
\item Differential photon flux $\varphi$ of the simulation at $E = 1\,$TeV in units of the Crab flux,
\item Photon index $\Gamma$ of the simulation,
\item Diffuse cone angle ($0^{\circ}$ for the simulation of a point source)
\item Number of computing processes $k$ per run.
\end{itemize}
The first three are \textit{run}-dependent and automatically read out from the DB, whereas the four latter are set by the user. Simulating a certain flux $\varphi$ instead of a fixed number of events per run implies a consistent statistics level throughout the production version 
and corresponds to the simulation of a constant source. 

MC simulations were carried out to determine up to which impact distances, corresponding to the distance of the shower axis prolonged to the ground with respect to the center of the array, electromagnetic showers can trigger the H.E.S.S. telescopes. Based on these, the maximum simulated impact distance is set to
\begin{equation}
R_{\mathrm{sim}}\left( \theta_{\mathrm{max}} \right) = R_{0} \times \frac{1}{\cos \theta_{\mathrm{max}}},
\label{Eq_Impact}
\end{equation} 
where $R_{0} = 550\,\mathrm{m}$. Both the value and the functional relation are notably almost independent of other observation parameters (other than the energy), which was verified with the underlying simulations. 
Another relevant input for the simulation is the minimum primary photon energy of a \textit{run}, which was similarly determined to
\begin{equation}
E_{\mathrm{min}} = \frac{E_{0}}{\varepsilon_{\mathrm{max}}} \times \frac{1}{\cos^3 \theta_{\mathrm{min}}},
\label{Eq_Energy}
\end{equation}
with $E_{0} = 5\,\mathrm{GeV}$ for \textit{runs} where the large H.E.S.S. telescope CT5 (see \cite{2014_CT5}) took part in the observation and $E_{0} = 30\,\mathrm{GeV}$ for all other \textit{runs}. $\varepsilon_{\mathrm{max}}$ corresponds to the maximum of the relative optical efficiencies of the observing IACTs, or directly to the one of CT5, if participating. The zenith-angle dependence in Eq.~\ref{Eq_Energy} takes into account both the fact that at larger $\theta$ the photon density on ground is lower due to the enlarged Cherenkov cone (see Eq.~\ref{Eq_Impact}) as well as the increased attenuation due to the larger optical depth.
The maximum energy $E_{\mathrm{max}}$ of the simulated primary photons is in our case fixed to $100\,\mathrm{TeV}$ for all observations, but can be adjusted if desired. 

As stated above, the user sets the photon index of the simulation and the differential flux $\varphi$ at $E = 1\,\mathrm{TeV}$. The latter is set in units of the Crab, where the reference spectrum of the Crab Nebula is taken from \cite{2015_MAGICCrab}. Together with $R_{\mathrm{sim}}$, $E_{\mathrm{min}}$, and $E_{\mathrm{max}}$, the simulation phase space for a given \textit{run} is set, allowing to derive the number of events $N$ to be simulated.
Depending on the intended application of the simulation, the parameters $\varphi$, $\Gamma$, and potentially $k$, are adjusted by the user. To facilitate finding appropriate parameters, the statistics can be evaluated via the interface prior to the simulation. The interface estimates the energy $E_{\mathrm{IRF,max}}$ up to which IRFs generated from the respective simulations are expected to be valid as follows: In the analysis framework at hand, IRFs are filled by grouping the simulated events that pass the analysis cuts ($N_{\gamma}$) in bins of width $\Delta E_{\mathrm{true}} / E_{\mathrm{true}} = 10\%$. For a proper calculation of all characteristics of interest, a minimum of $N_{\gamma} = 50$ is required per bin. Assuming that $N_{\mathrm{sim}} / N_{\gamma} \approx 10$ and since in the RWS approach IRFs are calculated on a \textit{run}-wise basis (see Section~\ref{spectra}), $E_{\mathrm{IRF,max}}$ corresponds to the energy where $N_{\mathrm{sim}} \left( \left[ E , 1.1 \times E \right] \right) > 500$ is still valid. In addition to $E_{\mathrm{IRF,max}}$, the approximate, combined computation time $t_{\mathrm{CPU}}$ for the shower and detector simulation is computed, which is assumed to be roughly proportional to the integrated energy of the primary photons:
\begin{equation}
t_{\mathrm{CPU}} = a \times \int_{E_{\mathrm{min}}}^{E_{\mathrm{max}}} E \times \frac{\mathrm{d}N}{\mathrm{d}E}\times \mathrm{d}E~.
\label{Eq_tcpu}
\end{equation}
It has to be noted that here $\mathrm{d}N / \mathrm{d}E$ already denotes the differential energy distribution of all particles simulated for a \textit{run} and is thus given in units of $1 / \mathrm{TeV}$. Test simulations have shown that generally $a < 10\,\mathrm{s} / \mathrm{TeV}$, and therefore $a = 10\,\mathrm{s} / \mathrm{TeV}$ is used for the computation to provide a conservative estimate of $t_{\mathrm{CPU}}$. All computation times refer to the HEP-SPEC06 (HS06) benchmarking standard. 
Actual computation times on recent CPUs are typically a factor of $10$ smaller. 

\begin{table}
\caption{Overview of the parameters of the two exemplary simulation scenarios $A$ and $B$. The derived event numbers and computation times refer to the simulation of one \textit{run}. CPU times refer to the HS06 standard; actual computation times on current machines are at least 10 times faster.}
\label{tab_simscen}
\begin{tabular}{lcc} \toprule
 & Scenario $A$ & Scenario $B$ \\  \midrule
$\varphi$ (Crab) & $10$ & $100$ \\
$\Gamma$ & $2.4$ & $1.8$ \\
$N_{\mathrm{CT1-4}}$ & $2.7\times 10^{5}$ & $8.4\times 10^{5}$ \\
$N_{\mathrm{CT5}}$ & $3.3\times 10^{6}$ & $3.5\times 10^{6}$ \\
$E_{\mathrm{IRF,max}}$ (TeV) & $1.2$ & $24$ \\
$t_{\mathrm{CPU,CT1-4}}$ (hr) & $140$ & $1900$ \\ 
$t_{\mathrm{CPU,CT1-5}}$ (hr) & $300$ & $2000$ \\ 
 \bottomrule
\end{tabular}
\end{table}

The outlined parameter calculation is illustrated for two exemplary scenarios in Table~\ref{tab_simscen}. Both scenarios correspond to $\theta = 30^{\circ}$, $\varepsilon_{\mathrm{max}} = 80\%$, and a \textit{run} duration of $28\,$min. In case of a Crab-like source (or fainter), scenario $A$ provides at least a factor of $10$ more simulated than measured gamma-rays over the complete energy range. The corresponding statistics level is thus sufficient for general MC studies as well as the extraction of the $1$D point spread function (PSF, see Sec.~\ref{morphology}).
For a different spectral shape of the source, $\Gamma$ should be adjusted accordingly. The derived values of $t_{\mathrm{CPU}}$ for this case are rather moderate, allowing to easily generate RWS of typical data sets for the full H.E.S.S. array on a sufficiently large computing grid. Viewed differently, the simulation of newly taken data of the five-telescope H.E.S.S. array for scenario $A$ in the case of $1000\,$h of observation per year requires at least $8$ continuously operating computing cores of the current generation (again assuming a scaling factor of 10 with respect to the HS06 standard).

While in principle scenario $A$ provides appropriate event statistics for a variety of use cases, generating simulations with a spectral energy distribution similar to the source of interest generally implies only few events at higher energies. If the RWS are intended to be used for the generation of IRF tables, a minimum number of gamma-like events per IRF energy bin are required as outlined above. The easiest solution for better high-energy event statistics would be to just raise $\varphi$ in scenario $A$, but has the disadvantage of overall higher statistics throughout all simulated energies, which is not needed. The parameters of scenario $B$ are thus chosen such as to increase high-energy statistics without simulating too many photons at lower energies. In this particular case, the IRFs are usable up to at least $E_{\mathrm{IRF,max}} = 24\,$TeV, which can be even more increased by varying $\varphi$ and $\Gamma$. Generating simulations for this scenario requires more computation time, but is still feasible, especially with computing facilities such as the EGI, to which the H.E.S.S. collaboration has access.

\subsection{Observation-Based Simulation Configuration}
\label{sec_simconf}

The RWS framework exploits a large variety of information of each observation \textit{run} to enhance the description of the simulation, as is going to be laid out in the following. As mentioned in Section~\ref{sec_techframe}, this information is stored in the DB and read out for the simulation.

\begin{table*}\centering
\caption{List of parameters that are used for the simulation of a \textit{run}.}
\label{tab_simpar}
\begin{tabular}{lcc}
\toprule
Quantity & Context & Comment \\
                           & (per pixel, telescope, or for array) & \\  \midrule
active IACTs & - & - \\
start and end time & array & see section \ref{sec_simconf} \\
telescope tracking & array & see section \ref{sec_techframe} \\
source position & array & see section \ref{sec_simconf} \\
optical efficiency $\varepsilon$ & telescope  & see \cite{2006_HessCrab} \\
Transparency coefficient & array & see section \ref{sec_simconf} and \cite{2014_TC}  \\
camera focus & telescope & only relevant for CT5; see \cite{2014_CT5} \\ 
trigger settings & telescope & see \cite{2006_HessCrab} \\
live-time fraction & telescope & see section \ref{sec_simconf} and \cite{Trigger} \\
broken pixels & pixel & for High and Low Gain; see \cite{Balzer2010aga}\\
PMT gain & pixel & see \cite{2004_HESSI_Cameras} \\
Hi-Lo ratio (HG/LG) & pixel & see \cite{2004_HESSI_Cameras} \\
Flatfield coefficient (FF) & pixel & see \cite{2004_HESSI_Cameras} \\
Night-Sky Background & pixel & see section \ref{sec_simconf} \\
 \bottomrule
\end{tabular}
\end{table*}
An overview of the \textit{run} information that is used for the simulation is given in Table~\ref{tab_simpar}. The correct use of all input parameters has been thoroughly and successfully checked. To prepare the shower simulation, the start and end times of a given \textit{run} are read out from the DB. The $N$ events to be simulated are equally distributed over the time window, and a unique time stamp is assigned to each event. In case the simulation of a \textit{run} is split into more than one computing process, the time window of the \textit{run} is split accordingly. The input coordinates to be used for the simulation are given in J2000 sky coordinates. Two sets of coordinates are set and fixed throughout the \textit{run}: The pointing position, corresponding to the observation position of the telescopes during the corresponding real data taking, and the putative source position. This allows simulating sources that are not centred in the field of view, as commonly performed by IACTs during the so-called ``wobble'' observation mode. The source position can either be set manually or read out automatically for \textit{runs} where a specific source was observed. If diffuse simulations are to be generated, a diffuse cone angle around the source position is set additionally, and the directions of the simulated particles are distributed uniformly within this circle. Although the resulting event distribution is homogeneous within the cone angle, such simulations are still suitable for generating arbitrary source morphologies through event rescaling at analysis level (see Section~\ref{morphology}).

At the beginning of the simulation of each particle, the pointing direction of the telescopes is updated to simulate the apparent trajectory of the source on the sky. The directional origin of the particle is converted accordingly. This approach corresponds to a complete, realistic simulation of the apparent movement of a constant source in the night sky which is being tracked by an array of IACTs. The accuracy of the telescope tracking is mainly limited by pointing uncertainties, for which a reliable \textit{run}-wise measurement is however complicated \cite{2010_Hess_GC_Position}. They are thus mimicked via a random but constant pointing offset of each IACT throughout the \textit{run}, determined from a 2D Gaussian with $\sigma = 40''$. This value is inferable from the systematic pointing error of H.E.S.S. of $20''$ per axis \citep{2004_Gillessen} when assumed to be the result of independent telescope-wise offsets for a system of 4 identical IACTs. 

To account for variations of the atmospheric quality, the Cherenkov transparency coefficient (TC) \cite{2014_TC} is read out from the DB. The TC is multiplied to the simulated relative optical efficiency of each participating IACT, providing a first-order correction as compared to the standard approach. An alternative way of exploiting the TC information would be to directly modify the photon attenuation during the shower simulation process but has not yet been implemented. An even more accurate treatment of atmospheric dependencies is to directly use a refined atmospheric profile for the simulation, for example one obtained from Lidar measurements during the same run as shown by \cite{2019_LidarRWS}. 
Another advancement of RWS is the exploitation of the live-time fraction of each telescope according to which it is randomly set unavailable for a given event because of the readout dead time. This leads to a more realistic distribution of event multiplicities with a corresponding influence on, e.g., the PSF. While this can in principle also be incorporated in classical simulation frameworks, the live-time fraction of each telescope depends on various circumstances, most importantly the array and camera configuration. 

The RWS framework furthermore simulates the major variable photomultiplier tube (PMT) characteristics (see Table~\ref{tab_simpar}), i.e. operates on a pixel-wise basis. Though these can in general exhibit intra-\textit{run} variations, all of them are fixed for each \textit{run} in the current implementation. It is obvious that the quality of the resulting simulations strongly depends on a working calibration framework, which serves as the source of most of the parameters.
\begin{figure}[]
\includegraphics[width=\columnwidth]{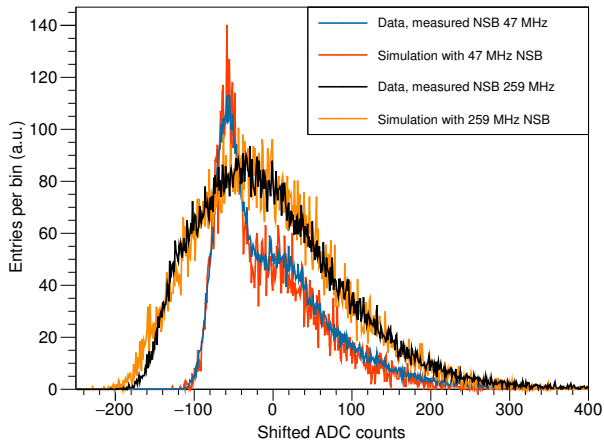}
\caption{Measured and simulated pedestal shapes of two specific pixels with different NSB levels as a function of the charge (centered around $0$).}
\label{Fig_Pedestal}
\end{figure}
As an example, the overall agreement of the measured and simulated pedestal shapes of two given pixels with different level of night-sky background (NSB) are shown in Figure~\ref{Fig_Pedestal}. This agreement is the result of a fully working calibration and simulation chain: For each pixel, the rate of the NSB is determined from the respective pedestal width obtained from the data \textit{run} and stored in the DB during regular data taking. For the RWS, this value as well as the gain of the respective PMT is again read out from the DB, and NSB photons are simulated using that rate. Whereas for \textit{standard} MC simulations the NSB level is generally set constant throughout the FoV, even very inhomogeneous NSB patterns are correctly simulated in the new approach.

\subsection{RWS MC Analysis}
\label{mc_ana}

A set of RWS \textit{runs} can be analysed in the same manner as actual data. A list of previously simulated \textit{runs} is passed to the high-level analysis framework, and a test position is chosen in sky coordinates. This usually corresponds to the position of the simulated gamma-rays, which is in most cases identical to the catalogue position of the source of interest. For the MC analysis, the simulations can be re-weighed to a user-defined spectral shape. Commonly used shapes, such as power-law, log-parabola, and exponential cut-off power-law are already implemented, and any needed shape could easily be added. The user chooses between using the full available event statistics or alternatively a differential flux to rescale to. The rescaling and reshaping is implemented such that either a corresponding weighing factor is assigned to a given event (\textit{event weighing}), or that this factor is used as a probability to randomly keep the event (\textit{event throwing}). Similarly, a temporal event rescaling mechanism at analysis level would allow simulating variable sources but has not yet been implemented.  
The output of the analysis is a file which is given in the same format as the results file of the analysis of real data. In addition to the information about the analysis configuration, it contains distributions of cut parameters, reconstruction variables etc., allowing for detailed comparisons of simulation and data. As a proper reconstructed sky coordinate is assigned to each event, even the generation of sky-coordinate-based plots is possible.

\section{Applications and Performance}
\label{sec_perf}

The RWS framework brings a variety of different use cases and advancements, some of which are going to be discussed in this section.

\subsection{Comparison of Simulation and Data}

The analysis of realistically simulated gamma-rays from RWS of a given data set has the advantage of delivering an appropriate reference to the excess events of the actual data, provided that the systematic uncertainties of the background subtraction are sufficiently well under control. In principle, a basic comparison of, e.g., the distribution of analysis cut variables is also possible with classical MC simulations. However the proper combination of observation and detector configurations under the assumption of a specific spectral shape as described in Sec.~\ref{mc_ana} allows a more thorough assessment including that of systematic errors.
\begin{figure*}[]
\includegraphics[width=\textwidth]{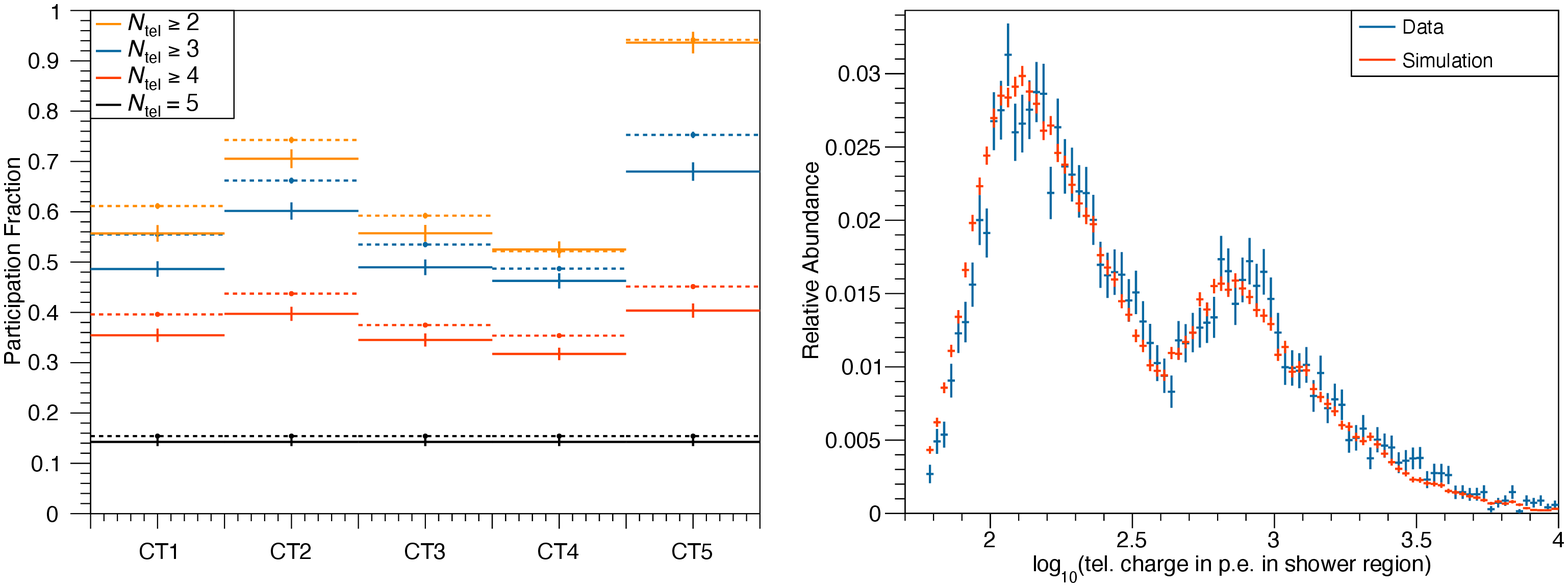}%
\caption{Exemplary RWS-MC vs. data comparisons for a CT1-5 data set of the Crab Nebula. \textit{Left:} Distribution of relative telescope participation. Dashed lines denote the output from RWS and solid ones the actual data. \textit{Right:} Relative abundance of the decadic logarithm of telescope-wise intensities in the respective shower region. The cutoff on the left side is a direct result of the reconstruction as only events with $\ge 60\,$p.e. are kept.}
\label{Fig_TelPart}
\end{figure*}
An example for such a comparison of RWS-MC simulations and data is given on the \textit{left panel} of Fig.~\ref{Fig_TelPart}. It shows the relative participation of each telescope to events with different number of triggered telescopes $N_{\mathrm{tel}}$ for a given data set taken on the Crab Nebula with CT1-5 and reconstructed and analysed with the \textit{Model} analysis \citep{2009_Model} in stereoscopic mode. The energy spectrum used for the event reshaping is the one from \citep{2015_MAGICCrab}. The resulting values of the distribution are in this case strongly influenced by parameters like the active IACTs, their optical efficiency $\varepsilon$, the zenith angle $\theta$, and the assumed energy spectrum of the source, requiring dedicated effort to perform a MC-data comparison of this distribution with classical simulations for a large data set. With the RWS framework however, all these characteristics are implicitly handled and match the observed distributions quite well. Visible discrepancies, such as for $N_{\mathrm{tel}} \ge 3$, are the result of systematic uncertainties like the one on the spectral shape, which can be studied with distributions like the present one. Another example of such a comparison is shown for the same data set and analysis on the \textit{right panel} of Fig.~\ref{Fig_TelPart}, depicting the relative abundance of the decimal logarithm of the telescope-wise intensities in the respective shower region. While at first sight the peculiar double-peaked structure might hint at, e.g., analysis problems with the data, the fact that the shape is well reproduced by the simulations is reassuring. After such a validation an explanation for the observed behaviour comes more naturally. In this case it is a result of the stereoscopic analysis of telescopes with different sizes; only events where at least two telescopes were triggered are considered in this analysis, and for these the average light yield of CT5 is higher than for the others, leading to the peak on the right. Given that the charge of each telescope with data is individually added to this plot, the left peak turns out to be higher because of events with $N_{\mathrm{tel}} > 2$.
The combined study of such parameter distribution comparisons helps in identifying problems and assessing the general agreement of data and simulation.

\subsection{Influence of Observation/Detector Conditions}

As laid out in Section~\ref{sec_simconf}, various information regarding the observation and detector configuration during a \textit{run} enters the RWS framework, expecting to render the resulting simulations more realistic than the ones generated with pre-defined settings. 
For a given parameter the quantitative improvement of this approach mostly depends on the parameter itself as well as the discrepancy of its value used for classical simulations. This implies that a corresponding assessment is generally only valid for the investigated data set and analysis case, and given that RWS are conceptually different from classical simulations a comparison of the two does not allow to disentangle the influences of individual parameters. A better way of studying the imprint of a specific condition is the comparison of RWS with the nominal \textit{run} settings with a second one where the settings of a given parameter are modified. An input parameter for which this approach works well is the NSB. For classical H.E.S.S. simulations it is usually set to $100\,$MHz per pixel throughout the FoV, which is thus used for comparison in this study. The exemplary \textit{run} investigated here was taken with CT1-4, pointed to a position with about $1.2^{\circ}$ angular distance to the Galactic Centre.
\begin{figure*}[]
\includegraphics[width=\textwidth]{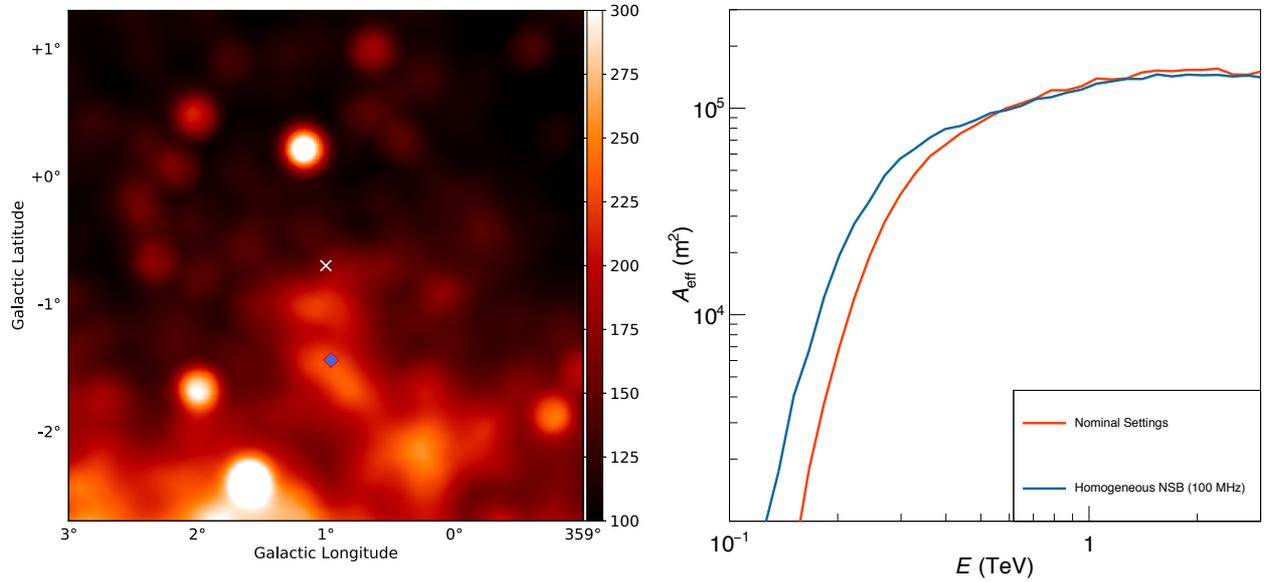}
\caption{\textit{Left:} NSB map of a region near the Galactic Centre. The pointing position of the \textit{run} studied here is highlighted with a white cross, and the simulation position used to derive the effective areas on the \textit{right} panel is shown with a blue diamond. The values on the colour bar are given in units of MHz per pixel.  \textit{Right:} Analysis effective area of the position on the \textit{left} for a specific observation \textit{run} taken with CT1-4. The red curve corresponds to the result of the full RWS MC, whereas for the blue one the NSB level was set to $100\,$MHz per pixel throughout the complete FoV.}
\label{Fig_NSB}
\end{figure*}
This position is drawn on the \textit{left} panel of Fig.~\ref{Fig_NSB} on top of the NSB level of the region, reconstructed from the pixel pedestal widths as described in Section~\ref{sec_simconf}. As is typical in the vicinity of the Galactic Plane, the NSB rate is strongly position-dependent because of numerous localised emission regions, in this case ranging from around $100\,$MHz to more than $300\,$MHz per pixel. To test the corresponding influence on the analysis, point-like gamma-rays from a position with enhanced NSB level of around $220\,$MHz were simulated, once using the full \textit{run} information including the NSB pattern and a second time with homogenised $100\,$MHz per pixel but otherwise identical conditions. The resulting analysis effective areas, again using the \textit{Model} analysis \citep{2009_Model} with the \textit{Standard} configuration, are plotted on the \textit{right} panel of Fig.~\ref{Fig_NSB}. While they are almost inseparable at $E > 500\,$GeV, they differ quite severely below that. Even for a source with comparatively hard spectral index of $\Gamma = 2.2$ this results in $20\%$ difference in event rate over the complete energy range, which increases to almost $60\%$ for a soft-spectrum source with $\Gamma = 3.6$. For energies below $500\,$GeV, the difference even increases to $50\%$ ($\Gamma = 2.2$) respectively $80\%$ ($\Gamma = 3.6$). It is important to emphasise that these numbers are only valid for this specific setup and may not be generalised, not even to sky regions with nominally similar NSB level. Although in the description above the NSB rate at the simulation position was mentioned, it has to be kept in mind that because of the angular offset of source direction and Cherenkov shower in the camera focal plane the difference of the analysis effective areas is rather the result of the overall NSB pattern \textit{around} the test position. This complexity also implies that, although simulations with different homogeneous NSB levels can be generated and interpolated with the classical simulation approach, only RWS provide a correct treatment of this decisive parameter for regions like the one shown here.
Another aspect to be highlighted is that the influence of, e.g., the NSB on analysis characteristics also strongly depends on the applied reconstruction method and event selection. The response to such influences is usually the result of an involved interplay at different stages, underlining the importance of simulations that are as realistic as possible.

\subsection{Spectral Fitting}
\label{spectra}

Spectral fit results obtained with IRFs from RWS are expected to be more precise because of smaller systematic uncertainties. The IRF generation with RWS MCs as well as the corresponding spectrum and light curve calculation have been adapted to the software framework. IRFs are filled and read out on a per-run basis, hence again no interpolation of lookup tables is necessary.

\begin{figure*}[]
\includegraphics[width=\textwidth]{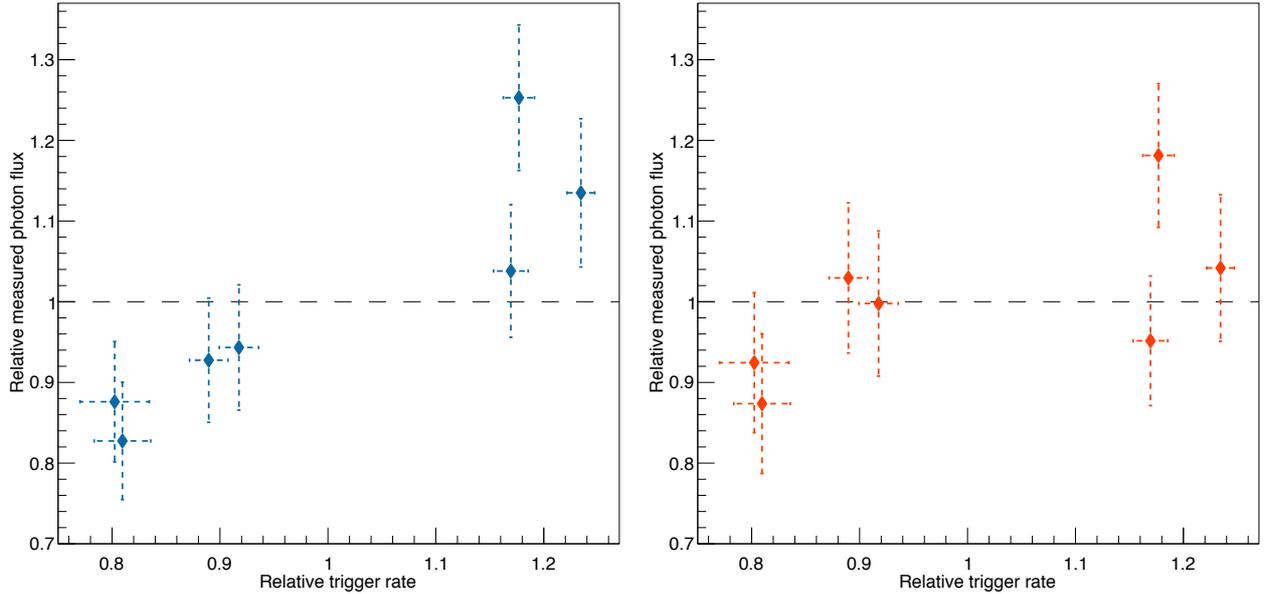}
\caption{Dependence of the gamma-ray flux on the trigger rate (relative to the mean value for each axis) for IRFs from classical (\textit{left panel}) and \textit{run}-wise simulations (\textit{right panel}), using the same Crab Nebula data set.}
\label{Fig_LCCorr}
\end{figure*}
The influence of the used IRF type on the resulting light curve is illustrated with a small data set, consisting of seven \textit{runs} taken with CT1-4 on the Crab Nebula. The reconstructed photon flux of each \textit{run}, normalised to the respective mean to only show relative differences, is plotted against the event trigger rate in Fig.~\ref{Fig_LCCorr}. When using IRFs from classical simulations, the photon flux appears directly correlated with the trigger rate, and the compatibility with a constant flux is quite poor ($\chi^2 / \textrm{NDoF} = 2.86$). Main responsible factors are changing weather and atmospheric conditions, interpolation of IRFs at different zenith angles and non-functional camera pixels. As most \textit{run}-specific conditions are included in the RWS and the resulting IRFs, that tendency is considerably less pronounced in this case, and the flux is for this data set basically constant within statistical uncertainties ($\chi^2 / \textrm{NDoF} = 1.10$). It has to be noted that this test was deliberately conducted with a reduced data set as it is only meant to demonstrate the capabilities of the RWS approach and does not correspond to a detailed evaluation of the underlying systematic uncertainties of both approaches. 

\subsection{Morphology Fitting}
\label{morphology}

Providing a more realistic simulation approach, RWS are very well suited for the extraction of the instrument PSF, thus allowing more precise morphological studies.

With the classical simulation approach, lookup tables are often generated to store the PSF for different varying observation configurations. When using RWS, the PSF of a given source can be directly simulated by analysing the simulated data set (see Section~\ref{mc_ana}). Depending on the use case, the PSF can be generated and extracted in different ways. To obtain the one-dimensional angular resolution under the assumption of radial symmetry, typically a $\vartheta^2$ histogram (where $\vartheta$ denotes the angular distance of reconstructed and true source position) with a fine binning is extracted. This resulting MC $\vartheta^2$ histogram is then used as the reference PSF for the morphology fitting. 
\begin{figure}[]
\includegraphics[width=\columnwidth]{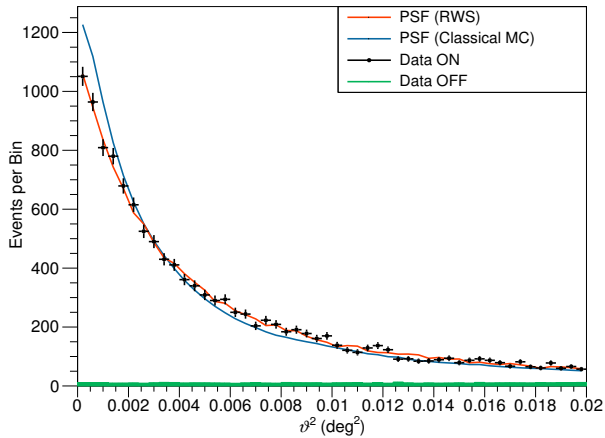}
\caption{Comparison of the $\vartheta^2$ distributions for a CT1-4 data set of the AGN Markarian 421. The distribution from the actual data is shown in black, the corresponding estimated background level is drawn in green. The red and blue lines denote the expected distributions of a point-like source as obtained from RWS and classical MC simulations, respectively.}
\label{Fig_Theta2}
\end{figure}
An example for such a PSF histogram is given in Fig.~\ref{Fig_Theta2} for a CT1-4 data set taken on the Active Galactic Nucleus (AGN) Markarian 421. With a declination of more than $38^{\circ}$, it is observable for H.E.S.S. only at very large zenith angles ($\theta > 60^{\circ}$), thus posing a strong challenge for the simulation framework. When using the PSF as obtained from classical MC simulations, there is a considerable mismatch between simulation and data, which could either be explained by PSF mis-modeling due to systematic uncertainties or by an intrinsic source extension. While the latter can for an individual source not be ruled out, such a mismatch between PSF and data has been systematically observed for basically all tested AGN, unambiguously pointing towards a systematic origin.
The RWS PSF on the other hand describes the measured distribution for Markarian 421 exceptionally well despite the complicated observation conditions, meaning that the source looks point-like for H.E.S.S. Knowing the PSF to better accuracy is boosting the power of statistical tests conducted with it. This is e.g. the case for the discrimination of competing spatial models of one single source, allowing to resolve smaller structures, or the separation of a composition of multiple sources.
The first application of this new potential was the detection and measurement of the extension of the Crab Nebula in gamma-rays \citep{2019_CrabHess}.

While the approach of extracting the PSF from the simulated $\vartheta^2$ histograms also works well in most cases for two-dimensional morphology studies, it relies on the assumption that the PSF is radially symmetric. In reality however, the PSF at a given position in the FoV is a two-dimensional function which is influenced by the analysis settings as well as the overall observation conditions of the data set. Parameters such as the array configuration can distort the PSF and additionally introduce an overall shift of the reconstructed direction. Since most respective influences are included in RWS, it is possible to prevent these problems by directly extracting the two-dimensional PSF in sky coordinates from the results file of the MC analysis.
\begin{figure*}[]
\includegraphics[width=\textwidth]{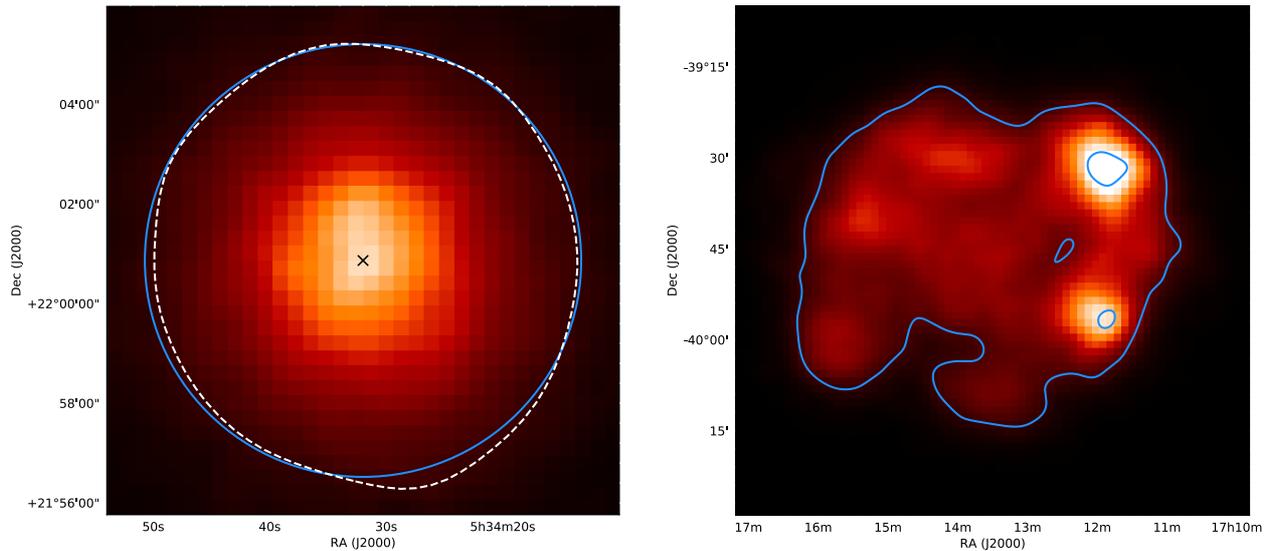}%
\caption{Exemplary applications of RWS for morphological studies on different scales. \textit{Left:} Reconstructed directions of simulated gamma-rays from a point source for \textit{runs} with only three participating IACTs in sky coordinates. The corresponding $68\%$ event containment contour is shown in dashed white. For comparison, the blue circle illustrates the radially symmetric $68\%$ containment radius, centered on the origin of the simulated gamma-rays. The latter, corresponding to the position of the Crab pulsar, is drawn as a black cross. \textit{Right:} Simulated, background-free H.E.S.S. counts map of RX J1713.7-3946, using the XMM-Newton map (\cite{HESS_RXJ1713}, contours denoting $5\%$ and $50\%$ of the maximum intensity in blue) as a template.}
\label{Fig_MorphRWS}
\end{figure*}
An example is given on the \textit{left} panel of Fig.~\ref{Fig_MorphRWS}, showing the simulated H.E.S.S. PSF of the Crab Nebula for \textit{runs} where only three out of the four smaller H.E.S.S. telescopes took data, making the active part of the IACT array asymmetric.
The two-dimensional $68\%$ containment radius significantly deviates from radial symmetry and is seen to be slightly offset with respect to the simulation position. Using the two-dimensional PSF from RWS for the morphology fit thus implies more reliable results, allowing to test more complex source models for extensions near the resolution limit. Furthermore the reconstructed source position is generally expected to be more precise, which becomes especially important for sources near the FoV edge. The two afore-mentioned approaches correspond to complementary ways to extract the angular detector response to point-like sources. As for morphology studies of extended sources, another method has been developed in the RWS framework of the H.E.S.S. software. In case a prediction of the source morphology is available (for example from a model or from multi-wavelength observations), it can be used as a template for the MC analysis. As a first step of such an analysis, the template is rescaled such that the maximum value in the map is $1$. Using diffuse simulations with a sufficiently large cone angle, the simulated direction of each event is evaluated, and the corresponding value of the rescaled template is used either as an event weight or as a probability to keep it. The approach is the morphological analogy to the spectral re-shaping described in Section~\ref{mc_ana}. An example is given on the \textit{right} panel of Fig.~\ref{Fig_MorphRWS}, which shows the simulated H.E.S.S. map of RX J1713.7-3946, using the XMM-Newton map from \cite{HESS_RXJ1713} as a template. The observation conditions and offsets are implicitly handled in the resulting map, and it can thus be directly compared to the results from the actual data analysis.
This allows conducting multi-wavelength comparisons like the one in \cite{HESS_RXJ1713} with much more detail and a better PSF accuracy, and even has the potential to lower systematic uncertainties of derived fluxes by allowing to calculate more accurate IRFs in such cases.
\section{Conclusions and Prospects}
\label{sec_conc}

Here we introduced a novel simulation approach for IACT arrays which is designed to generate custom MC simulations for every observation \textit{run} of a data set, allowing to incorporate observation and detector conditions at a greater detail. The method has been fully developed and implemented in one of the simulation and analysis chains of H.E.S.S., from the automatic simulation parameter calculation up to the adaptation of the high-level analysis modules. As the concept leads to more realistic simulations, it not only implies a reduction but also allows a better assessment of systematic errors, opening up new science cases. Enabling a new level of precision in terms of angular resolution, it played a key role in the first measurement of extension at the arcminute level of the Crab Nebula at TeV energies \cite{2019_CrabHess}.
Regarding spectral measurements, the better sensitivity regarding lightcurve measurements will help for detection of source variability. By combining the spectral and morphological advancements discussed in Sections~\ref{spectra} and \ref{morphology}, IRFs can in principle be calculated for any given source shape assumption, being more precise than the current approach of using the same lookup tables for all extended sources.

The more accurate treatment of factors such as an inhomogeneous NSB level across the FoV are furthermore expected to help improve the background modelling, which is crucial for, e.g., measuring diffuse VHE emission at large scales. It should be noted that the aforementioned influences, such as NSB, are expected to have a greater effect for next-generation IACT arrays because of their larger FoV. Besides reduced systematic uncertainties, the RWS approach has the advantage that the necessary computational resources are independent of the simulated complexity or the amount of different observation configurations, scaling linearly with the observation time of the corresponding data. The RWS concept as introduced in this publication therefore becomes increasingly attractive for complex IACT arrays such as the upcoming Cherenkov Telescope Array \cite{2019_ScienceWithCTA} as it may be the only computationally viable way to reach the desired level of systematics, helping to push the achievable science output to a higher level.

\section*{Acknowledgements}

We thank the H.E.S.S. Collaboration for allowing us to use the data in this publication. We are grateful to C. van Eldik, W. Hofmann, as well as the two anonymous referees for carefully reading the manuscript and providing us with very useful suggestions. Finally, our thanks go to all the members of the H.E.S.S. Collaboration for their technical support and for many stimulating discussions. This work was supported by the French ANR (project ANR-13-IS05-0001) as well as the Austrian FWF (I1345-N27) science funds.

\bibliographystyle{elsarticle-num}
\addcontentsline{toc}{part}{Bibliography}

\begin{thebibliography}{10}
\expandafter\ifx\csname url\endcsname\relax
  \def\url#1{\texttt{#1}}\fi
\expandafter\ifx\csname urlprefix\endcsname\relax\def\urlprefix{URL }\fi
\expandafter\ifx\csname href\endcsname\relax
  \def\href#1#2{#2} \def\path#1{#1}\fi

\bibitem{2016_HessUpgrade}
{G.~Giavitto et al.}, {Upgraded cameras for the HESS imaging atmospheric
  Cherenkov telescopes}, in: Society of Photo-Optical Instrumentation Engineers
  (SPIE) Conference Series, Vol. 9908 of SPIE Proceedings, 2016.
\newblock \href {http://dx.doi.org/10.1117/12.2231865}
  {\path{doi:10.1117/12.2231865}}.

\bibitem{2011_VeritasUpgrade}
A.~{Nepomuk Otte}, {for the VERITAS Collaboration}, {The Upgrade of VERITAS
  with High Efficiency Photomultipliers}, Proceedings of the 32nd ICRC,
  Beijing.

\bibitem{2016_MagicUpgrade}
{J.~Aleksi{\'c} et al.}, {The major upgrade of the MAGIC telescopes, Part I:
  The hardware improvements and the commissioning of the system}, Astroparticle
  Physics 72 (2016) 61--75.
\newblock \href {http://dx.doi.org/10.1016/j.astropartphys.2015.04.004}
  {\path{doi:10.1016/j.astropartphys.2015.04.004}}.

\bibitem{2009_Model}
M.~{de Naurois}, L.~{Rolland}, {A high performance likelihood reconstruction of
  {$\gamma$}-rays for imaging atmospheric Cherenkov telescopes}, Astroparticle
  Physics 32 (2009) 231--252.
\newblock \href {http://dx.doi.org/10.1016/j.astropartphys.2009.09.001}
  {\path{doi:10.1016/j.astropartphys.2009.09.001}}.

\bibitem{2014_ImPACT}
R.~D. {Parsons}, J.~A. {Hinton}, {A Monte Carlo template based analysis for
  air-Cherenkov arrays}, Astroparticle Physics 56 (2014) 26--34.
\newblock \href {http://dx.doi.org/10.1016/j.astropartphys.2014.03.002}
  {\path{doi:10.1016/j.astropartphys.2014.03.002}}.

\bibitem{2009_TMVA}
S.~{Ohm}, C.~{van Eldik}, K.~{Egberts}, {{$\gamma$}/hadron separation in
  very-high-energy {$\gamma$}-ray astronomy using a multivariate analysis
  method}, Astroparticle Physics 31 (2009) 383--391.
\newblock \href {http://dx.doi.org/10.1016/j.astropartphys.2009.04.001}
  {\path{doi:10.1016/j.astropartphys.2009.04.001}}.

\bibitem{2006_HessCrab}
{F.~{Aharonian} et al.}, {Observations of the Crab nebula with HESS}, Astronomy
  and Astrophysics 457 (2006) 899--915.
\newblock \href {http://dx.doi.org/10.1051/0004-6361:20065351}
  {\path{doi:10.1051/0004-6361:20065351}}.

\bibitem{2016_MAGICCrab}
{J.~Aleksi{\'c} et al.}, {The major upgrade of the MAGIC telescopes, Part II: A
  performance study using observations of the Crab Nebula}, Astroparticle
  Physics 72 (2016) 76--94.
\newblock \href {http://dx.doi.org/10.1016/j.astropartphys.2015.02.005}
  {\path{doi:10.1016/j.astropartphys.2015.02.005}}.

\bibitem{2013_CTA_MC}
{K.~Bernl{\"o}hr et al.}, {Monte Carlo design studies for the Cherenkov
  Telescope Array}, Astroparticle Physics 43 (2013) 171--188.
\newblock \href {http://dx.doi.org/10.1016/j.astropartphys.2012.10.002}
  {\path{doi:10.1016/j.astropartphys.2012.10.002}}.

\bibitem{geant4_2003}
{S.~Agostinelli et al.}, {Geant4 - A Simulation Toolkit}, Nuclear Instruments
  and Methods in Physics Research Section A: Accelerators, Spectrometers,
  Detectors and Associated Equipment 506~(3) (2003) 250--303.
\newblock \href {http://dx.doi.org/10.1016/S0168-9002(03)01368-8}
  {\path{doi:10.1016/S0168-9002(03)01368-8}}.

\bibitem{1994_Kaskade}
M.~P. {Kertzman}, G.~H. {Sembroski}, {Computer simulation methods for
  investigating the detection characteristics of TeV air Cherenkov telescopes},
  Nuclear Instruments and Methods in Physics Research A 343 (1994) 629--643.
\newblock \href {http://dx.doi.org/10.1016/0168-9002(94)90247-X}
  {\path{doi:10.1016/0168-9002(94)90247-X}}.

\bibitem{1998_Corsika}
{D.~Heck et al.}, {CORSIKA: a Monte Carlo code to simulate extensive air
  showers}, 1998.

\bibitem{2003_Thesis_Guy}
J.~Guy, \href{tel.archives-ouvertes.fr/tel-00003488}{{Premiers r{\'e}sultats de
  l'exp{\'e}rience HESS et {\'e}tude du potentiel de d{\'e}tection de
  mati{\`e}re noire supersym{\'e}trique}}, Phd thesis, {Universit{\'e} Pierre
  et Marie Curie - Paris VI} (May 2003).
\newline\urlprefix\url{tel.archives-ouvertes.fr/tel-00003488}

\bibitem{2014_CT5}
{J.~Bolmont et al.}, {The camera of the fifth H.E.S.S. telescope. Part I:
  System description}, Nuclear Instruments and Methods in Physics Research A
  761 (2014) 46--57.
\newblock \href {http://dx.doi.org/10.1016/j.nima.2014.05.093}
  {\path{doi:10.1016/j.nima.2014.05.093}}.

\bibitem{2015_MAGICCrab}
{J.~Aleksi{\'c} et al.}, {Measurement of the Crab Nebula spectrum over three
  decades in energy with the MAGIC telescopes}, Journal of High Energy
  Astrophysics 5 (2015) 30--38.
\newblock \href {http://dx.doi.org/10.1016/j.jheap.2015.01.002}
  {\path{doi:10.1016/j.jheap.2015.01.002}}.

\bibitem{2014_TC}
{J.~Hahn et al.}, {Impact of aerosols and adverse atmospheric conditions on the
  data quality for spectral analysis of the H.E.S.S. telescopes}, Astroparticle
  Physics 54 (2014) 25--32.
\newblock \href {http://dx.doi.org/10.1016/j.astropartphys.2013.10.003}
  {\path{doi:10.1016/j.astropartphys.2013.10.003}}.

\bibitem{Trigger}
{S.~{Funk}, et al.}, {The trigger system of the H.E.S.S. telescope array},
  Astroparticle Physics 22~(3-4) (2004) 285--296.
\newblock \href {http://arxiv.org/abs/astro-ph/0408375}
  {\path{arXiv:astro-ph/0408375}}, \href
  {http://dx.doi.org/10.1016/j.astropartphys.2004.08.001}
  {\path{doi:10.1016/j.astropartphys.2004.08.001}}.

\bibitem{Balzer2010aga}
A.~Balzer, {Systematic studies of the H.E.S.S. camera calibration}, Ph.D.
  thesis, U. Erlangen-Nuremberg (main) (2010).

\bibitem{2004_HESSI_Cameras}
{F.~{Aharonian} et al.}, {Calibration of cameras of the H.E.S.S. detector},
  Astroparticle Physics 22 (2004) 109--125.
\newblock \href {http://dx.doi.org/10.1016/j.astropartphys.2004.06.006}
  {\path{doi:10.1016/j.astropartphys.2004.06.006}}.

\bibitem{2010_Hess_GC_Position}
{H.E.S.S.~ Collaboration}, {Localizing the VHE {$\gamma$}-ray source at the
  Galactic Centre}, MNRAS 402 (2010) 1877--1882.
\newblock \href {http://dx.doi.org/10.1111/j.1365-2966.2009.16014.x}
  {\path{doi:10.1111/j.1365-2966.2009.16014.x}}.

\bibitem{2004_Gillessen}
S.~{Gillessen},
  \href{www.bsz-bw.de/cgi-bin/xvms.cgi?SWB11244050}{Sub-{B}ogenminuten-genaue
  {P}ositionen von {T}e{V}-{Q}uellen mit {H}.{E}.{S}.{S}.}, Dissertation,
  Ruprecht-Karls-Universit{\"a}t Heidelberg, Heidelberg, Germany (2004).
\newline\urlprefix\url{www.bsz-bw.de/cgi-bin/xvms.cgi?SWB11244050}

\bibitem{2019_LidarRWS}
{J.~{Devin} et al.}, {Impact of H.E.S.S. Lidar profiles on Crab Nebula data},
  Vol. 197 of European Physical Journal Web of Conferences, 2019.
\newblock \href {http://dx.doi.org/10.1051/epjconf/201919701001}
  {\path{doi:10.1051/epjconf/201919701001}}.

\bibitem{2019_CrabHess}
{H.E.S.S.~ Collaboration}, {Resolving the Crab pulsar wind nebula at
  teraelectronvolt energies}, Nature Astronomy (2019) 476\href
  {http://dx.doi.org/10.1038/s41550-019-0910-0}
  {\path{doi:10.1038/s41550-019-0910-0}}.

\bibitem{HESS_RXJ1713}
{H.E.S.S.~ Collaboration}, {H.E.S.S. observations of RX J1713.7-3946 with
  improved angular and spectral resolution: Evidence for gamma-ray emission
  extending beyond the X-ray emitting shell}, Astronomy and Astrophysics 612
  (2018) A6.
\newblock \href {http://dx.doi.org/10.1051/0004-6361/201629790}
  {\path{doi:10.1051/0004-6361/201629790}}.

\bibitem{2019_ScienceWithCTA}
{The CTA Consortium}, {Science with the Cherenkov Telescope Array}, World
  Scientific, 2019.

\end{thebibliography}

\end{document}